\documentclass[twocolumn,floatfix,aps,prl,showpacs,superscriptaddress,preprintnumbers,amsmath,amssymb]{revtex4}

\usepackage{graphicx}
\usepackage{bm}

\def\Journal#1#2#3#4{{#1}{\bf #2}, #3 (#4)}

\def\NIMA{{Nucl. Instr. Meth.}~{\bf A}}

\def\PLB{{Phys. Lett.}~{\bf B}}
\def\PLC{Phys. Repts.\ }
\def\PRL{Phys. Rev. Lett.\ }
\def\PRD{{Phys. Rev.}~{\bf D}}

\begin{document}

\title{ $J/\psi$ production from proton-proton collisions at  
$\sqrt{s}$~=~200~GeV }

\newcommand{\abilene}{Abilene Christian University, Abilene, TX 79699, USA}
\newcommand{\acadsin}{Institute of Physics, Academia Sinica, Taipei 11529, Taiwan}
\newcommand{\banaras}{Department of Physics, Banaras Hindu University, Varanasi 221005, India}
\newcommand{\barc}{Bhabha Atomic Research Centre, Bombay 400 085, India}
\newcommand{\bnl}{Brookhaven National Laboratory, Upton, NY 11973-5000, USA}
\newcommand{\caucr}{University of California - Riverside, Riverside, CA 92521, USA}
\newcommand{\ciae}{China Institute of Atomic Energy (CIAE), Beijing, People\'s Republic of China}
\newcommand{\cns}{Center for Nuclear Study, Graduate School of Science, University of Tokyo, 7-3-1 Hongo, Bunkyo, Tokyo 113-0033, Japan}
\newcommand{\columbia}{Columbia University, New York, NY 10027 and Nevis Laboratories, Irvington, NY 10533, USA}
\newcommand{\dapnia}{Dapnia, CEA Saclay, F-91191, Gif-sur-Yvette, France}
\newcommand{\debrecen}{Debrecen University, H-4010 Debrecen, Egyetem t{\'e}r 1, Hungary}
\newcommand{\fsu}{Florida State University, Tallahassee, FL 32306, USA}
\newcommand{\gsu}{Georgia State University, Atlanta, GA 30303, USA}
\newcommand{\hiroshima}{Hiroshima University, Kagamiyama, Higashi-Hiroshima 739-8526, Japan}
\newcommand{\ihepprot}{Institute for High Energy Physics (IHEP), Protvino, Russia}
\newcommand{\isu}{Iowa State University, Ames, IA 50011, USA}
\newcommand{\jinrdubna}{Joint Institute for Nuclear Research, 141980 Dubna, Moscow Region, Russia}
\newcommand{\kaeri}{KAERI, Cyclotron Application Laboratory, Seoul, South Korea}
\newcommand{\kangnung}{Kangnung National University, Kangnung 210-702, South Korea}
\newcommand{\kek}{KEK, High Energy Accelerator Research Organization, Tsukuba-shi, Ibaraki-ken 305-0801, Japan}
\newcommand{\kfki}{KFKI Research Institute for Particle and Nuclear Physics (RMKI), H-1525 Budapest 114, POBox 49, Hungary}
\newcommand{\korea}{Korea University, Seoul, 136-701, Korea}
\newcommand{\kurchatov}{Russian Research Center ``Kurchatov Institute", Moscow, Russia}
\newcommand{\kyoto}{Kyoto University, Kyoto 606, Japan}
\newcommand{\labllr}{Laboratoire Leprince-Ringuet, Ecole Polytechnique, CNRS-IN2P3, Route de Saclay, F-91128, Palaiseau, France}
\newcommand{\lawllnl}{Lawrence Livermore National Laboratory, Livermore, CA 94550, USA}
\newcommand{\losalamos}{Los Alamos National Laboratory, Los Alamos, NM 87545, USA}
\newcommand{\lpc}{LPC, Universit{\'e} Blaise Pascal, CNRS-IN2P3, Clermont-Fd, 63177 Aubiere Cedex, France}
\newcommand{\lund}{Department of Physics, Lund University, Box 118, SE-221 00 Lund, Sweden}
\newcommand{\muenster}{Institut fuer Kernphysik, University of Muenster, D-48149 Muenster, Germany}
\newcommand{\myongji}{Myongji University, Yongin, Kyonggido 449-728, Korea}
\newcommand{\nagasaki}{Nagasaki Institute of Applied Science, Nagasaki-shi, Nagasaki 851-0193, Japan}
\newcommand{\newmex}{University of New Mexico, Albuquerque, NM, USA}
\newcommand{\nmsu}{New Mexico State University, Las Cruces, NM 88003, USA}
\newcommand{\ornl}{Oak Ridge National Laboratory, Oak Ridge, TN 37831, USA}
\newcommand{\orsay}{IPN-Orsay, Universite Paris Sud, CNRS-IN2P3, BP1, F-91406, Orsay, France}
\newcommand{\pnpi}{PNPI, Petersburg Nuclear Physics Institute, Gatchina, Russia}
\newcommand{\riken}{RIKEN (The Institute of Physical and Chemical Research), Wako, Saitama 351-0198, JAPAN}
\newcommand{\rkrbrc}{RIKEN BNL Research Center, Brookhaven National Laboratory, Upton, NY 11973-5000, USA}
\newcommand{\saispbstu}{St. Petersburg State Technical University, St. Petersburg, Russia}
\newcommand{\saopaulo}{Universidade de S{\~a}o Paulo, Instituto de F\'{\i}sica, Caixa Postal 66318, S{\~a}o Paulo CEP05315-970, Brazil}
\newcommand{\seoulnat}{System Electronics Laboratory, Seoul National University, Seoul, South Korea}
\newcommand{\stonybrkc}{Chemistry Department, Stony Brook University, SUNY, Stony Brook, NY 11794-3400, USA}
\newcommand{\stonycrkp}{Department of Physics and Astronomy, Stony Brook University, SUNY, Stony Brook, NY 11794, USA}
\newcommand{\subatech}{SUBATECH (Ecole des Mines de Nantes, CNRS-IN2P3, Universit{\'e} de Nantes) BP 20722 - 44307, Nantes, France}
\newcommand{\tenn}{University of Tennessee, Knoxville, TN 37996, USA}
\newcommand{\titech}{Department of Physics, Tokyo Institute of Technology, Tokyo, 152-8551, Japan}
\newcommand{\tsukuba}{Institute of Physics, University of Tsukuba, Tsukuba, Ibaraki 305, Japan}
\newcommand{\vandy}{Vanderbilt University, Nashville, TN 37235, USA}
\newcommand{\waseda}{Waseda University, Advanced Research Institute for Science and Engineering, 17 Kikui-cho, Shinjuku-ku, Tokyo 162-0044, Japan}
\newcommand{\weizmann}{Weizmann Institute, Rehovot 76100, Israel}
\newcommand{\yonsei}{Yonsei University, IPAP, Seoul 120-749, Korea}
\affiliation{\abilene}
\affiliation{\acadsin}
\affiliation{\banaras}
\affiliation{\barc}
\affiliation{\bnl}
\affiliation{\caucr}
\affiliation{\ciae}
\affiliation{\cns}
\affiliation{\columbia}
\affiliation{\dapnia}
\affiliation{\debrecen}
\affiliation{\fsu}
\affiliation{\gsu}
\affiliation{\hiroshima}
\affiliation{\ihepprot}
\affiliation{\isu}
\affiliation{\jinrdubna}
\affiliation{\kaeri}
\affiliation{\kangnung}
\affiliation{\kek}
\affiliation{\kfki}
\affiliation{\korea}
\affiliation{\kurchatov}
\affiliation{\kyoto}
\affiliation{\labllr}
\affiliation{\lawllnl}
\affiliation{\losalamos}
\affiliation{\lpc}
\affiliation{\lund}
\affiliation{\muenster}
\affiliation{\myongji}
\affiliation{\nagasaki}
\affiliation{\newmex}
\affiliation{\nmsu}
\affiliation{\ornl}
\affiliation{\orsay}
\affiliation{\pnpi}
\affiliation{\riken}
\affiliation{\rkrbrc}
\affiliation{\saispbstu}
\affiliation{\saopaulo}
\affiliation{\seoulnat}
\affiliation{\stonybrkc}
\affiliation{\stonycrkp}
\affiliation{\subatech}
\affiliation{\tenn}
\affiliation{\titech}
\affiliation{\tsukuba}
\affiliation{\vandy}
\affiliation{\waseda}
\affiliation{\weizmann}
\affiliation{\yonsei}
\author{S.S.~Adler}	\affiliation{\bnl}
\author{S.~Afanasiev}	\affiliation{\jinrdubna}
\author{C.~Aidala}	\affiliation{\bnl}
\author{N.N.~Ajitanand}	\affiliation{\stonybrkc}
\author{Y.~Akiba}	\affiliation{\kek} \affiliation{\riken}
\author{J.~Alexander}	\affiliation{\stonybrkc}
\author{R.~Amirikas}	\affiliation{\fsu}
\author{L.~Aphecetche}	\affiliation{\subatech}
\author{S.H.~Aronson}	\affiliation{\bnl}
\author{R.~Averbeck}	\affiliation{\stonycrkp}
% ...
\collaboration{PHENIX Collaboration} \noaffiliation

\date{\today}        
\begin{abstract}

$J/\psi$ production has been measured in proton-proton collisions at 
$\sqrt{s}$~=~200~GeV over a wide rapidity and transverse momentum range by the 
PHENIX experiment at RHIC. Distributions of the rapidity and transverse 
momentum, along with measurements of the mean transverse momentum and total 
production cross section are presented and compared to available theoretical 
calculations. The total $J/\psi$ cross section is 3.99 $\pm$ 0.61(stat) 
$\pm$ 0.58(sys) $\pm$ 0.40(abs) $\mu$b. The mean transverse momentum is 1.80 
$\pm$ 0.23(stat) $\pm$ 0.16(sys) GeV/c.

\end{abstract}
\maketitle



Understanding $J/\psi$ production mechanisms requires data over a large 
range of collision energies and with broad coverage in rapidity and 
transverse momentum ($p_T$). Existing data at lower energies from fixed target 
hadron experiments yield total cross sections and mean $p_T$ 
(${\langle}p_T{\rangle}$) values in the energy range $\sqrt{s}$ = 
7~-~38.8~GeV~\cite{fixed_target}. Limited kinematic coverage in collider 
experiments~\cite{fixed_target,cdfd0_jpsi,cdf_prompt_jpsi} has 
so far meant that total cross sections and mean $p_T$ values
could not be measured. The systematic study of 
$J/\psi$ production at Relativistic Heavy Ion Collider (RHIC) energies 
with wide $p_T$ and rapidity coverage should therefore provide crucial 
tests of $J/\psi$ production models. In addition, the RHIC 
proton-proton results provide a baseline for studying cold and hot nuclear 
matter in proton-nucleus and nucleus-nucleus collisions using $J/\psi$ 
yields as a probe.

Intense theoretical interest in the $J/\psi$ production mechanism was 
stimulated when the Color Singlet Model (CSM) was found~\cite{csm} to 
dramatically 
underpredict the high $p_T$ CDF prompt $J/\psi$ and $\psi(2S)$ cross 
sections~\cite{cdf_prompt_jpsi}. Attention turned toward models in which 
color octet $c\bar{c}$ states can also contribute to the $J/\psi$ yield. The 
Color Octet Model (COM), which is based on the Non-Relativistic QCD 
model~\cite{cdf_nrqcd}, has been successful in reproducing the 
high $p_T$ CDF prompt $J/\psi$ cross sections, as has the more 
phenomenological Color Evaporation Model~\cite{cem}. 

In this paper we report results of the first measurements of 
$pp \rightarrow J/\psi + X$ at RHIC, made at $\sqrt{s}$ = 200~GeV by the 
PHENIX experiment. The data yield the first total cross sections 
for $J/\psi$ production beyond fixed target energies, and the first 
measurement of ${\langle}p_T{\rangle}$
beyond $\sqrt{s}$ = 63 GeV. They will constrain models in the lower $p_T$ 
region where gluon fusion is expected to dominate (at $p_T$ beyond about 5 
GeV/c, the direct $J/\psi$ production cross section is expected to be dominated
by fragmentation of high $p_T$ gluons~\cite{fragmentation}).


The PHENIX experiment~\cite{phenix_nim} detects electrons in the 
pseudo-rapidity range ${\mid}\eta{\mid} \le 0.35$ in two central 
spectrometer arms covering $\Delta\phi$~=~90$^{\circ}$, and forward rapidity 
muons in two muon arms covering $\Delta\phi$~=~360$^{\circ}$. Only one muon arm, 
covering $1.2 < \eta < 2.2$, was operational for this data set. Electrons are 
identified by matching charged particle tracks to energy 
deposits in the Electromagnetic Calorimeter (EMC) and to rings in 
the Ring Imaging {\v C}erenkov Detector (RICH), which has a threshold of 
4.7 GeV/c for pions. Muons are identified by finding deeply penetrating 
roads in the Muon Identifier (MuID) and matching them to tracks in 
the Muon Tracker (MuTr).

The data were recorded during the 2001/2002 $pp$ run at $\sqrt{s}$ = 200 
GeV. After quality assurance and vertex cuts ($\pm$ 35 cm for $ee$ and 
$\pm$ 38 cm for $\mu\mu$), 67 nb$^{-1}$ were used for the $J/\psi \rightarrow 
\mu^+ \mu^-$ analysis, and 82 nb$^{-1}$ for $J/\psi 
\rightarrow e^+ e^-$. The minimum bias interaction trigger required at least
one hit on each side of the interaction vertex in the Beam-Beam counter 
(BBC). Minimum bias trigger rates varied from 5~to~30 kHz. Events containing 
$J/\psi$ decays were selected using level-1 
triggers in coincidence with the minimum bias interaction trigger. 
The $J/\psi \rightarrow e^+e^-$ trigger required a minimum energy deposit of 
either 0.75~GeV in a 2~$\times$~2 tile of EMC towers or 2.1~GeV in a
4~$\times$~4 tile. The $J/\psi \rightarrow \mu^+ \mu^-$ trigger required 
at least two deeply penetrating roads in separate azimuthal quadrants of 
the MuID~\cite{hiroki_thesis}.


$J/\psi$ yields in the central arms were obtained by reconstructing 
electron-positron pairs. Electron candidates were charged particle 
tracks that were associated with a RICH ring ($\ge 2$ hit phototubes)
and an EMC hit ($\pm 4 \sigma$ position association cut), and which 
satisfied $0.5 < E/p < 1.5$, where $E$ is the EMC cluster energy and 
$p$ is the reconstructed track momentum. A 5 GeV/c upper limit on 
electron momentum prevented charged pions from firing the RICH. 


$J/\psi$ yields in the muon arm were obtained by reconstructing 
$\mu^+\mu^-$ pairs. Muon tracks were reconstructed by finding a 
track seed in the MuID and matching it to clusters of hits in each 
of the three MuTr stations. The momentum was determined by fitting, 
with a correction for energy loss, the MuID and MuTr hit positions 
and the BBC vertex position. Each track was required to pass $5 \sigma$ 
cuts on the $\chi^2$ from the track fit and on the radial 
distance of the fitted track from the measured $z$-vertex position. 

\begin{figure}
\includegraphics[width=\linewidth]{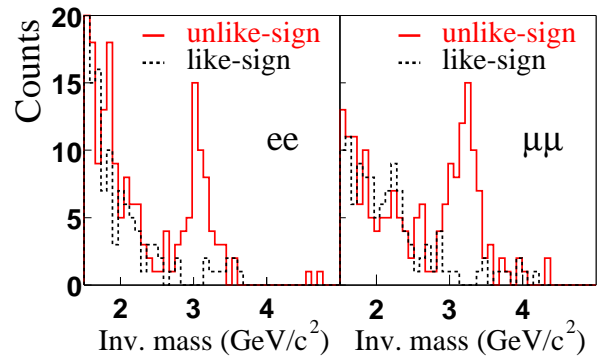}
\caption{\label{fig:inv_mass}
The invariant mass spectra for dielectron and dimuon pairs. 
Unlike-sign pairs are shown as solid lines, the sum of like-sign pairs as
dashed lines.}
\end{figure}

Unlike-sign pairs and, for background estimation, like-sign 
pairs satisfying the above conditions were combined to form invariant 
mass spectra. Simulations show that the acceptance for like-sign and 
unlike-sign pairs is the same to within a few percent for invariant 
masses above 1~GeV/c$^2$ for electrons. In Fig.~\ref{fig:inv_mass}, 
unlike-sign and like-sign 
invariant mass spectra from the entire $pp$ data set are shown together. 
For electrons, the net yield  inside 2.8~-~3.4~GeV/c$^2$ is 46, for muons 
inside 2.71~-~3.67~GeV/c$^2$ it is 65. For electrons, the peak width is 
110 MeV/c$^2$ and the centroid agrees well with the PDG value~\cite{pdg}. 
For muons, the width is 160~MeV/c$^2$. The muon peak centroid is higher 
than the PDG value by about 3\%, consistent with the uncertainty in the 
muon magnetic field calibration.

The $J/\psi$ cross sections were determined from the measured yields using 

\begin{displaymath} 
B_{ll} 
\frac{ d^2\sigma_{J/\psi} }{ dy dp_T} \,=\,
\frac{ N_{J/\psi} }{ (\int{{\cal L}dt}) \, {\Delta}y \, {\Delta}p_T} \,
\frac{ 1 }{ \epsilon_{bias} \, \epsilon_{lvl1} } \,
\frac{ 1 } { A \, \epsilon_{rec} }
\label{eq:dsigdpt_alt}
\end{displaymath}
where $N_{J/\psi}$ is the measured $J/\psi$ yield, $\int{{\cal L}dt}$ is 
the integrated luminosity measured by the minimum bias trigger, $B_{ll}$ 
is the branching fraction for the $J/\psi$ to either $e^+e^-$ or $\mu^+\mu^-$ 
pairs (PDG average value 5.9\%~\cite{pdg}), $\epsilon_{bias}$ is the 
minimum bias 
trigger efficiency 
for an event containing a $J/\psi$, $\epsilon_{lvl1}$ is the level-1 trigger 
efficiency for detecting a $J/\psi$, and $A \, \epsilon_{rec}$ is the 
acceptance times reconstruction efficiency for a $J/\psi$.

The integrated luminosity can be written as $\int{{\cal L}dt} =  
N_{MB}/\sigma_{BBC}$, where $N_{MB}$ is the number of minimum bias
triggers and $\sigma_{BBC}$ is the minimum bias trigger cross section. 
Using a van der Meer scan measurement, $\sigma_{BBC}$ was determined to be 
21.8 $\pm$ 2.1 mb~\cite{pi0_trigger}. We have estimated $\epsilon_{bias}$ 
in two ways. First, the minimum bias trigger efficiency for $J/\psi$ events 
from a simulation study using {\sc Pythia}~\cite{pythia} (with the GRV94NLO 
parton 
distribution functions (PDFs)) was 0.74, with no $p_T$ dependence. 
Good agreement is observed in the $dN_{ch}/d\eta$ distribution between 
{\sc Pythia} simulations and measurements~\cite{pythia_charged} for events 
involving high $p_T$ $\pi^0$ production. Second, we measured~\cite{pi0_trigger} 
our minimum bias trigger efficiency for high $p_T$ $\pi^0$ production using 
events recorded with a high $p_T$ EMC trigger. The efficiency of 0.75 
$\pm$ 3\% is constant within uncertainties over the measured range of 
$1.5 < p_T < 9$~GeV/c. We chose to use our measured trigger efficiency 
from high $p_T$ $\pi^0$ events when calculating the $J/\psi$ cross section, 
assuming that the trigger efficiency is the same for both processes.


For the electron analysis, $A \, \epsilon_{rec} \,$ was determined as a
function of $p_T$ using a full GEANT simulation of single $J/\psi$ events
with flat distributions in $dN/dy$ (${\mid}y{\mid} < 0.6$), 
$p_T$ ($p_T < 6$ GeV/c) and collision 
vertex (${\mid}z{\mid} < 35$ cm). The GEANT
simulations were tuned to match real detector responses for single
electrons.  The reconstruction efficiency calculations used a typical dead
channel map. An average correction for run-to-run variations in detector
active area was determined from single electron yields.  An estimate of
the systematic uncertainty in $A~\epsilon_{rec}$ due to $z$~vertex
dependence of the acceptance, momentum resolution effects, the pair mass
cut, and electron identification cuts is given in
Table~\ref{tab:systematic_error}, along with the uncertainty in the yield
due to Drell-Yan and correlated charm decay contributions.  The efficiency
$\epsilon_{lvl1}$ of the level-1 $J/\psi$ triggers in the central arms was
determined as a function of $p_T$ by using a software trigger-emulator to
analyze simulated single $J/\psi$ events.  The results are shown in
Table~\ref{tab:systematic_error}.  The trigger emulator was tuned by
analyzing simulated single electrons and comparing with the real single
electron trigger efficiency.

\begin{table}
\caption{\label{tab:systematic_error} 
Table of quantities and their systematic error estimates. Ranges are given
for $p_T$ dependent quantities.  For the $\mu^+\mu^-$ case the values of
$A \epsilon_{rec}$ and $\epsilon_{lvl1}$ are combined. The absolute cross
section normalization uncertainty from $\epsilon_{bias}$ and 
$\int{{\cal L}dt}$ is kept separate and is labeled (abs).}
\begin{ruledtabular}
  \begin{tabular}{ccc} 
   Quantity  & $e^+e^-$ &  $\mu^+\mu^-$ \\ \hline
   Yield & $\pm$ 5\% & $\pm$ 5\% \\
   $A \epsilon_{rec}$ & 0.026-0.010 $\pm$ 13\% & 0.038 - 0.017 $\pm$ 13\% \\
   $\epsilon_{lvl1}$ & (2$\times$2) 0.87-0.90 $\pm$ 5\% &  \\
                     & (4$\times$4) 0.30-0.74 $\pm$ 36\% &   \\
   $\epsilon_{bias}$ & 0.75 $\pm$ 3\% & 0.75 $\pm$ 3\% \\
   $\int{{\cal L}dt}$ & 82 nb$^{-1}$ $\pm$ 9.6\% & 67 nb$^{-1}$ $\pm$ 9.6\% \\
   Total & $\pm$ 15\%(sys) $\pm$ 10\%(abs) & $\pm$ 14\%(sys) $\pm$10\%(abs) \\
  \end{tabular}
\end{ruledtabular}
\end{table}


For the muon arm, $A \, \epsilon_{rec} \, \epsilon_{lvl1}$ was determined as 
an average within each rapidity and $p_T$ bin, using a full GEANT simulation
with $J/\psi$ events generated by {\sc Pythia} (with GRV94LO PDFs). The 
{\sc Pythia} $J/\psi$ rapidity and $p_T$ distributions are very similar to 
those of the real data, so that bin averaging effects should be approximately 
accounted for by this procedure. The simulated events were reconstructed 
using the same reconstruction software and cuts as for the real data, assuming 
nominal detector efficiencies and typical realistic dead channel and dead
high-voltage maps. Each event had to pass the simulated dimuon trigger. 
The systematic error includes discrepancies between Monte Carlo and real 
detector response, run to run variations in the detector state, and 
uncertainties in the {\sc Pythia} distributions. The results, integrated 
over the rapidity range of the muon arm, are shown in 
Table~\ref{tab:systematic_error}, along with an estimate of 
the systematic error on the yield due to the background subtraction technique.
In both the electron and muon cases the $J/\psi$ polarization was assumed to 
be zero, since existing $J/\psi$ polarization measurements are consistent 
with zero at low $p_T$~\cite{cdf_pol}. The effect of the unknown 
$J/\psi$ polarization has not been included in the systematic error. 

\begin{figure}
\includegraphics[width=\linewidth]{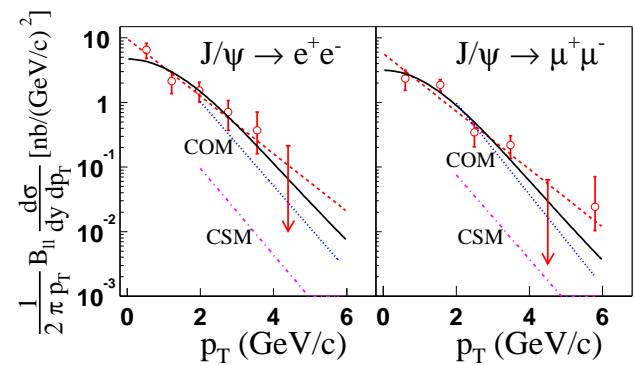}
\caption{\label{fig:combined}
The $J/\psi$ $p_T$ distributions for the dielectron and
dimuon measurements, with statistical uncertainties. The solid line 
is a phenomenological fit of the form 
$1/(2 \pi p_T)~d\sigma/dp_T~=~A~(1 + (p_T/B)^2)^{-6}$. The 
dashed line is an exponential fit. The CSM (dot-dashed)
and COM (dotted) calculations are from~\protect\cite{rhic_nrqcd}.} 
\end{figure}


The $p_T$ distributions for $J/\psi \rightarrow e^+e^-$ and $J/\psi 
\rightarrow \mu^+\mu^-$ are shown in Fig.~\ref{fig:combined}, with 
predictions~\cite{rhic_nrqcd} from the COM. Predictions of the 
CSM, which greatly underpredicts the cross sections, are also shown. These 
predictions are limited to $p_T>2$ GeV/c because parton intrinsic transverse 
momentum ($k_T$) broadening is not accounted for properly in the calculation. 
The COM calculations do not include fragmentation contributions, which become 
important at around 5 GeV/c~\cite{cdf_prompt_jpsi}. Calculations covering 
all $p_T$ and including fragmentation contributions are needed. The solid lines 
are a phenomenological fit of a form that has been shown to fit $J/\psi$ data 
well at fixed target energies~\cite{pt_phen}. The dashed line is an 
exponential fit. The phenomenological fits yield ${\langle}p_T{\rangle}$ 
values of 1.85 $\pm$ 0.46(stat) $\pm$ 0.16(sys) GeV/c (central arm) and 1.78 
$\pm$ 0.27(stat) $\pm$ 0.16(sys) GeV/c (muon arm), with a combined value of 
1.80 $\pm$ 0.23(stat) $\pm$ 0.16(sys) GeV/c. The systematic uncertainties were 
estimated from the spread in ${\langle}p_T{\rangle}$ from a weighted mean of 
the binned data, the phenomenological fit, and the exponential fit. An 
additional 3\% was assigned to the muon ${\langle}p_T{\rangle}$ due to the 
uncertainty in momentum scale.

\begin{figure}
\includegraphics[width=\linewidth]{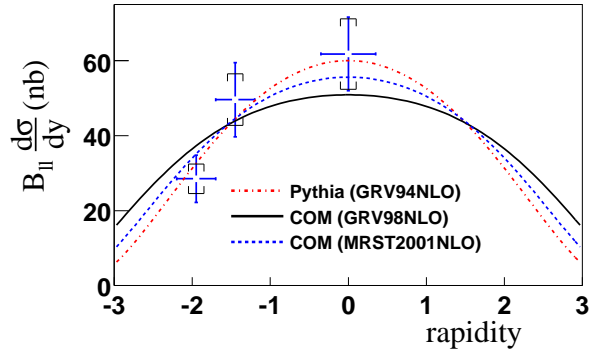}
\caption{\label{fig:rapidity_distribution}
The central rapidity point is from $J/\psi \rightarrow e^+e^-$, the 
others from $J/\psi \rightarrow \mu^+\mu^-$. The brackets represent systematic 
uncertainties. All curves have their overall normalization fitted to the data. 
The {\sc Pythia} shape was used to determine the cross section. There is an 
overall 10\% absolute normalization error not shown.} 
\end{figure}
\begin{figure}
\includegraphics[width=\linewidth]{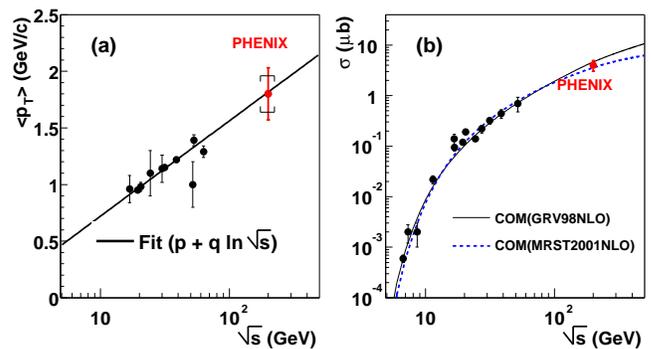}
\caption{\label{fig:sqrts}
(a) The present $J/\psi$ mean $p_T$ value compared with previous 
measurements at lower energy. The linear fit parameters are $p$ = 0.53, 
$q$ = 0.19. (b) The present $J/\psi$ total cross section compared with 
previous measurements at other values of $\sqrt{s}$. The curves are discussed 
in the text.}
\end{figure}

The $J/\psi$ rapidity distribution obtained by combining the dielectron and 
dimuon measurements is shown in Fig.~\ref{fig:rapidity_distribution}, with 
the muon arm data divided into two rapidity bins. 
The COM curves are theoretical shape predictions~\cite{hiroki_thesis} 
using the same models as are discussed in connection with Fig.~\ref{fig:sqrts}b, 
except that they  are normalized to our data to make the shape comparison 
clearer. Since gluon  fusion is the dominant process in all of the models, 
the rapidity shape depends mostly on the gluon distribution function and 
is not very sensitive to the production model. 
Most of the available PDFs are consistent with the data, 
and improved statistical precision will be needed to constrain them. 
A {\sc Pythia} calculation that reproduces the shape of our data best is 
also shown in Fig.~\ref{fig:rapidity_distribution}.  Normalizing this to 
the data, the total cross 
section was determined to be 3.99 $\pm$ 0.61(stat) $\pm$ 0.58(sys) $\pm$ 
0.40(abs)~$\mu$b. The quoted systematic error of 14\% was estimated by setting 
the measured cross sections all to their upper systematic error limits or all 
to their lower systematic error limits and noting how the cross section 
changed. The variation in the total cross section extracted if we use the same
procedure with different PDF choices and models was estimated to be 
small (~3\%).



A comparison is made in Fig.~\ref{fig:sqrts}a of the present 
${\langle}p_T{\rangle}$ value with values from previous 
experiments~\cite{fixed_target}. There are no theoretical predictions that 
we can compare with the ${\langle}p_T{\rangle}$ measurements. The total 
$J/\psi$ cross section determined in this analysis is shown in 
Fig.~\ref{fig:sqrts}b, along with cross sections determined by lower 
energy experiments~\cite{fixed_target} and predictions from the 
COM~\cite{hiroki_thesis} using two different PDFs. 
The $\sqrt{s}$ dependence of the cross section is sensitive to the 
factorization scale $Q$, since the shape of the PDFs depend on $Q$. 
The values of $Q$ (3.1 GeV for GRV98NLO and 2.3 GeV for MRST2001NLO) were 
chosen to give good agreement with the data. The total cross section 
normalization was obtained using color octet matrix elements 
from~\cite{beneke}, but has 
large theoretical uncertainties associated with the charm quark mass and the 
renormalization scale. The renormalization scale was taken to be equal to the 
quark mass M$_c$, and their values (1.48 GeV for GRV98NLO and 1.55 GeV for 
MRST2001NLO) were chosen to give good agreement with the data. The CEM is also 
able to describe the total cross section data~\cite{cem}. All measurements 
and models include feed-down from the $\chi_c$ and the $\psi'$ to the $J/\psi$.
We estimate~\cite{bfeed_down} that $B$ decay feed-down contributes less 
than 4\% to the $J/\psi$ total cross section at $\sqrt{s}$ = 200 GeV. 


In summary, we have presented the first $pp \rightarrow J/\psi + X$ 
measurements from RHIC, obtained at $\sqrt{s}$ = 200 GeV. The 
rapidity distributions, $p_T$ distributions, ${\langle}p_T{\rangle}$ and total 
cross sections have been presented and compared with available model 
calculations. The transverse momentum distributions above 2 GeV/c are 
reasonably well described by the COM. With the present statistical precision, 
our rapidity distribution shape is consistent with most of the available PDFs. 
COM calculations are able to reproduce the $\sqrt{s}$ dependence of the cross 
section using color octet matrix elements found in the literature, with a 
reasonable choice of QCD parameters. 


RHIC is expected to have proton-proton runs 
with enhanced luminosity at $\sqrt{s}$ = 200 and 500 GeV in the near 
future. The increased luminosity will improve the statistical 
precision and $p_T$ reach of the PHENIX data, and will ultimately make 
it possible to 
measure the $J/\psi$ polarization, which has been an important test for 
models~\cite{cdf_pol}.


We thank the staff of the Collider-Accelerator and Physics
Departments at BNL for their vital contributions.  We acknowledge
support from the Department of Energy and NSF (U.S.A.), MEXT and
JSPS (Japan), CNPq and FAPESP (Brazil), NSFC (China), CNRS-IN2P3
and CEA (France), BMBF, DAAD, and AvH (Germany), OTKA (Hungary), 
DAE and DST (India), ISF (Israel), KRF and CHEP (Korea),
RMIST, RAS, and RMAE, (Russia), VR and KAW (Sweden), U.S. CRDF 
for the FSU, US-Hungarian NSF-OTKA-MTA, and US-Israel BSF.

\end{document}